# Voltmeter in series?

Zoltan Gingl and Robert Mingesz

Department of Technical Informatics, University of Szeged, Árpád tér 2, 6720 Szeged, Hungary

E-mail: gingl@inf.u-szeged.hu

## Abstract
A recent physics challenge shows a circuit, where a voltmeter is connected in series. Indeed, real voltmeters have finite input resistance, therefore one may think that they can be used as resistors. In addition, voltmeters measure the voltage difference between their terminals, therefore it seems to be possible to calculate the current flowing through them. Is it okay? Does it make the voltmeter more universal? Are there any hidden secrets? How it is related to high-quality physics and STEM education, which are increasingly important in the modern world? Doesn't it approve an improper use that one can never see in any textbook and application? Suggesting and teaching such uncommon solutions doesn't generate undesired attitude? On the other hand, can it make the development of creativity and understanding harder if the students are taught to follow always the application rules? We do think it bears some discussion.

## The voltmeter-in-series circuit

Figure 1 shows the circuit of the challenge "*Why so series?*" [1, 2].

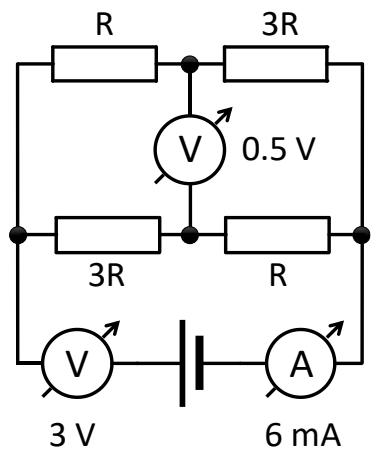

Figure 1. Circuit of the challenge [1, 2]. The meters show 6 mA, 0.5 V and 3.0 V.

Looking at the voltmeter in series can be confusing, since it is used as a conducting element. Although real voltmeters have finite input resistance and some current can flow through them, it is taught, that voltmeters should always be connected in parallel to the component on which the voltage drop is to be measured regardless of the value of their input resistance.

Following the anomalous use of the voltmeter shown above, one could connect a non-ideal ammeter in parallel and one could come up with a circuit, where two different voltage sources with non-zero internal resistances are connected in parallel, see figure 2. Looking at these circuits likely makes most of us uncomfortable, since we know that connecting voltage generators and ammeters in parallel can even cause damage.

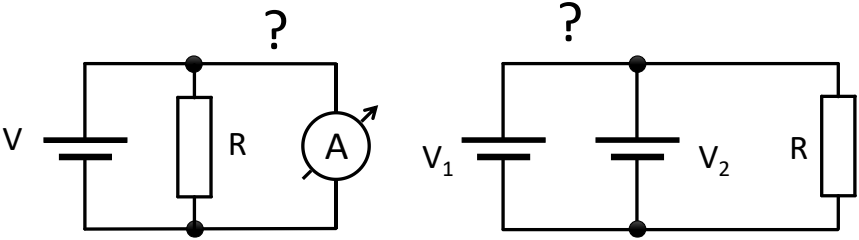

Figure 2. Theoretically it seems to be allowed to connect non-ideal ammeters and voltage sources with certain non-zero internal resistances in parallel.

Other examples can also be mentioned. Real capacitors can have some DC conductance (typically modelled by a resistor in parallel) and inductors can have considerable series resistance [3]. So, in theory, they could be used as resistors in DC circuits just like the voltmeter in figure 1. However, it is never done in practice.

We believe, it is worth to discuss the methodology behind the voltmeter-in-series example and especially its relation to education.

## Ideal and real voltmeters and ammeters

It is well known that a perfect instrument must not affect the operation of the observed system. Therefore, an ideal voltmeter behaves like an open circuit, while an ideal ammeter acts as a short circuit. In electronics, real components are modelled by combinations of ideal components. A real voltmeter is represented by an ideal voltmeter and a resistor, or more generally, an impedance, connected in parallel (see figure 3). So, it acts like an impedance, it can conduct current. Similar is true for real ammeters; they also function as impedances. From this point of view, only their very different practical impedance values separate them.

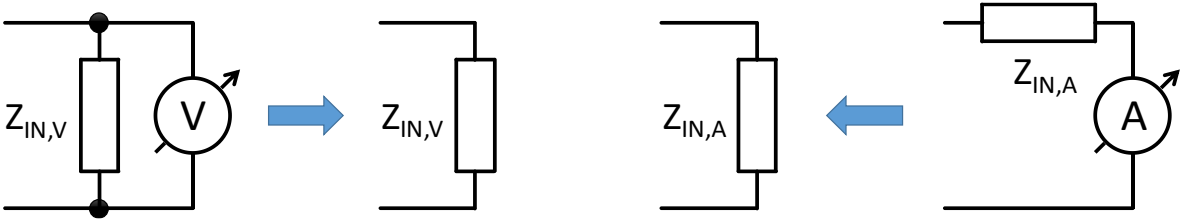

Figure 3. A real voltmeter or ammeter can be modelled as a parallel or serial combination of an ideal meter and an impedance, respectively. Both function as impedances in the circuit, but their real values are very different.

## Using voltmeters

In practice the impact of the input impedance of voltmeters on the system operation must be negligible. Therefore, there is always a certain limit of application, since the required accuracy must not be compromised by the presence of the input impedance. For most digital multimeters (DMM) its value is 10 MΩ, while oscilloscope inputs typically load the signal source with 1 MΩ. Note, that in the challenge the input resistance of the voltmeters equals 500 Ω, which is 20000 times less than the typical value for today's voltmeters.

In some exceptional cases low input impedance can be desirable. High speed signals may need proper termination to avoid reflections [4], therefore oscilloscope inputs can have a very low input impedance of 50 Ω. Low impedance (Lo Z) mode of voltmeters also exist to cancel the so-called ghost voltages typically present as a parasitic voltage in non-energized mains circuits [5]. Although the input impedance can be low, the voltmeters are always connected in parallel.

## Can the input impedance of a voltmeter be taken into account?

One can think that effect of the input impedance $Z_{IN}$ of the voltmeter can be taken into account and the voltmeter can even be used to measure the current flowing through it. For instance, the $V_G$ voltage of a generator connected to the voltmeter via a series resistor $R$ in figure 4 can be calculated as

$$V_G = \frac{R + Z_{IN}}{Z_{IN}} V \qquad (1)$$

where $V$ is the measured voltage.

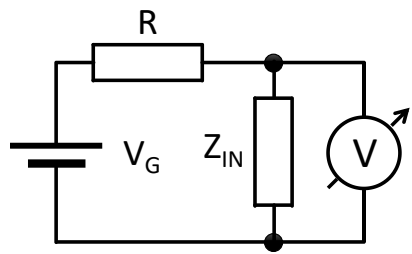

Figure 4. If the series resistor $R$ and the input impedance $Z_{IN}$ is known, $V_G$ can be determined using the value measured by the voltmeter.

Similarly, if a current $I$ flows through the voltmeter's input impedance (figure 5), its value can be expressed as

$$I = \frac{V}{Z_{IN}} \qquad (2)$$

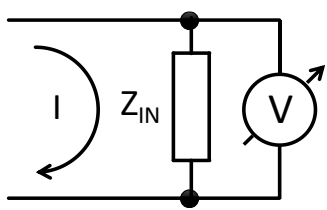

Figure 5. The voltmeter shows the voltage equal to the current multiplied by the input impedance.

For example, consider a 4.5 digit DMM having 10 MΩ input impedance according to the datasheet. If it displays a value of 1.0000 V and $R$ is 10 MΩ, then $V_G$ is calculated as 2.0000 V. In the same case, the current flowing through $Z_{IN}$ equals 0.1000 μA.

It seems to be very straightforward. However, doing so is not a good practice at all! Let us see why.

## Input impedance variants and specifications

All instruments have accuracy specifications and defined normal operating conditions, otherwise the measured value would be unreliable and useless. Although the voltage measurement accuracy of a voltmeter is always given, in contrast, the input impedance tolerance is rarely specified. Therefore, using the nominal value in calculations introduces unknown errors and can even cause the loss of reliability.

Let us see some examples. A leading manufacturer specifies the input impedance of their DMM only as >10 MΩ in DC voltage measurement mode (see the "Fluke 114, 115, 116 and 117 Digital Multimeters Extended specifications", https://dam-assets.fluke.com/s3fs-public/2793260_6116_ENG_A_W.PDF ). We think the message is clear: do not assume anything about its actual accuracy, use it only, when its effect is negligible. In other words, don't try to use the nominal value of the input impedance to compensate its effect or to calculate the current flowing through it. Note, that during AC voltage measurements the input impedance of the same instrument is given as >5 MΩ.

The manual of our UT60H 4.5 digit DMM used in education says that the voltage measurement accuracy is 0.1%+5 counts (means 0.15% overall error for 1.0000 V), while the input impedance is specified as "*Approx. 10 MΩ*". Some other DMMs' datasheet may say "*about 10 MΩ*", or "*10 MΩ nominal*" or just "*10 MΩ*". We have measured the input impedance of this DMM as a function of the input voltage, see figure 6.

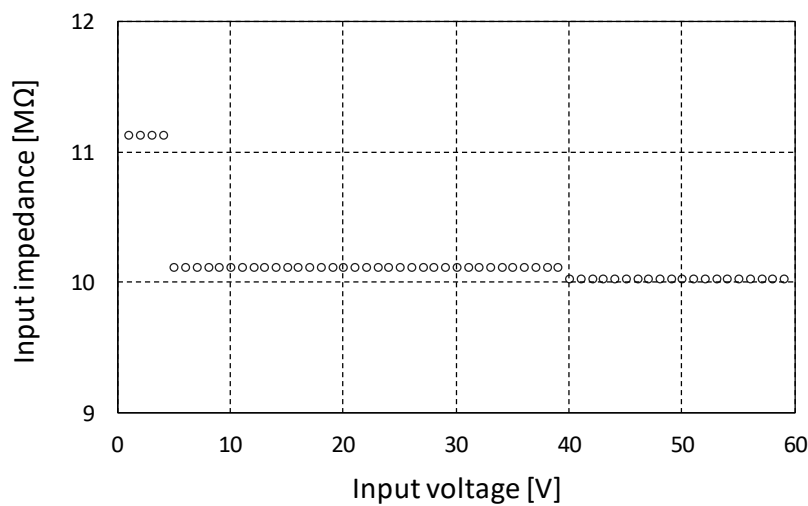

Figure 6. Input resistance as a function of the input voltage for a 4.5 digit multimeter under test (UT60H). The value is different in the measurement ranges of 4 V, 40 V and 400 V.

One can see that the value can be more than 10 % apart from the nominal value and it depends on the selected measurement range, too. So, using the measured 11.12 MΩ value in equation (1) and (2) gives 1.8993 V and 0.0899 µA in the abovementioned example instead of 2.0000 V and 0.1000 µA. Therefore, *using the nominal value introduces an error of about 33 and 67 times higher* than the error caused by the DMM itself!

What is the reason for such surprising and significant variations in the input impedance that can make even electronics enthusiasts and professionals sometimes confused?

There are typically two solutions in DMMs to set the input voltage range. Manual range selection is done by using a rotary switch that changes the voltage division ratio, see the left hand side circuit of figure 7. In this case the input voltage source always sees a constant load of close to 10 MΩ regardless of the range. Auto-ranging DMMs change the division ratio using integrated circuit switches, accordingly they typically employ a more suitable solution that can be seen on the right hand side of figure 7. The input impedance depends on the selected range in this case as we have observed. There are even such auto-ranging DMMs, for which the input impedance at the lowest ranges can be well above 1000 MΩ (all switches are off), while in other ranges their nominal impedance is still close to 10 MΩ.

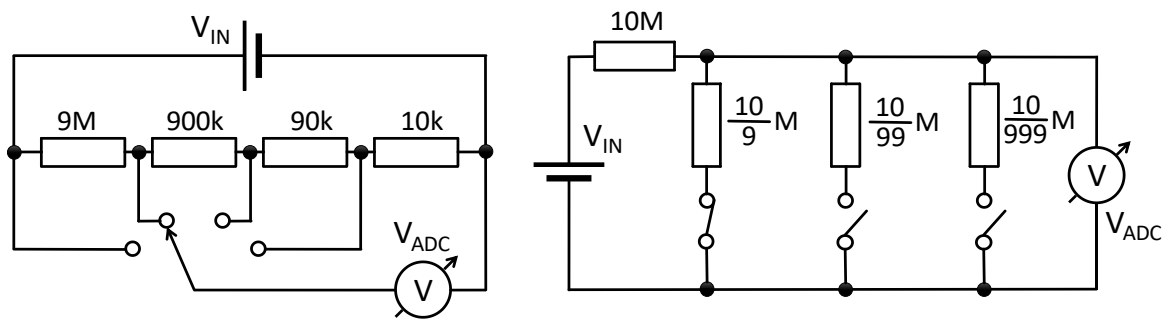

Figure 7. Principles of voltage divider circuits used in manual (left hand side) and auto-ranging voltmeters (right hand side) to select the division ratio of the input voltage $V_{IN}$ to provide a voltage $V_{ADC}$ for analogue-to-digital conversion.

In summary, there are several implementations of DMMs with various input impedance values that can even depend significantly on an automatically selected range. In general, the actual value and its accuracy can't be considered as known, in several cases only a minimum or approximate value is specified. Therefore, it is a bad practice to use the voltmeters in such conditions when their input impedance matters. There are only a few exceptions in some special cases. When using high-voltage DMM probes or attenuating oscilloscope probes, the input impedance can be a part of a voltage divider as shown in figure 4. However, in this case the user (typically a professional) must have an in-depth knowledge about the input impedance of the used instrument.

## Principles in STEM education

In technology, engineering and related sciences (but also in other fields including traffic, communication, economy, etc.) well-established standards and rules are used to guarantee proper operation and reliability which are essential for all electronic systems. Instruments, electronic components must always be used according to the specifications. In addition, there is no justifiable need to do the opposite. Violation of the rules and standards, ignoring implicit or explicit warnings can be dangerous, can lead to unexpected operation and getting used to it can result undesirable attitude [6,7]. High-quality STEM and related physics education play a very important role in developing carefulness, reliability and right attitude of professionals, engineers, scientists and teachers of the future. Education should prepare the students for real life jobs, therefore it should take well-established rules of technology and engineering seriously.

Let's go back to the circuit on figure 1 for a while. The intention likely was to recognize a really important feature of voltmeters. The problem is only with the message that connecting a voltmeter in series and applying it as a resistor is just normal, no reasons to worry, while teachers and students can never find this in any textbook or application. It certainly violates the rules of correct application. Teaching creative thinking about the input impedance can be done without such odd suggestion, see a recent paper in *Physics Education* [8]. The aim can be to observe the effect and to understand the limitations, but not to suggest irregular applications. One may say, that forcing to follow the rules limits the possibilities available for the students, however, it does not limit the development of creativity. It is a typical real-life challenge to find a smart solution when the conditions are given and engineers and scientists should be good in this.

Most of the teachers probably experience, that students often gain knowledge by memorizing figures, arrangements, they accept as recommended what they see during the lectures. Irregular use of a component without any warning can mislead students, but can be confusing even if the teacher says that it won't work properly in practice. Since understanding the operation out of normal conditions needs high level of expertise and care typically only experienced professionals, engineers and scientists have, such examples certainly exceed what can be expected from teachers and students.

## Conclusion

More and more modern tools and solutions are available for the students and teachers originally used only by experts including accurate electronic instruments, electronic components, sensors, microcontrollers (e.g. Arduino), software development tools, etc. These are very useful in high-quality STEM and physics education, but sometimes important rules of correct application are violated. We believe that tools, devices, components should never be used out of their normal operation modes. It is not needed to help understanding, furthermore it can be misleading, can generate unexpected and overlooked errors and can confirm questionable applications more generally [6,7,9]. Just like during teaching the traffic rules: never suggest to drive in the wrong direction in a one-way street or to go when the red light is on, even if it seems to be possible and safe. These are valid options only for authorized persons, like drivers of ambulance cars, and only in emergency. They must pass special training to learn what to do in such exceptional situations.

Finally, see figure 8 that draws the attention to the importance of teaching about the most important rules that helps students do the things right and safe.

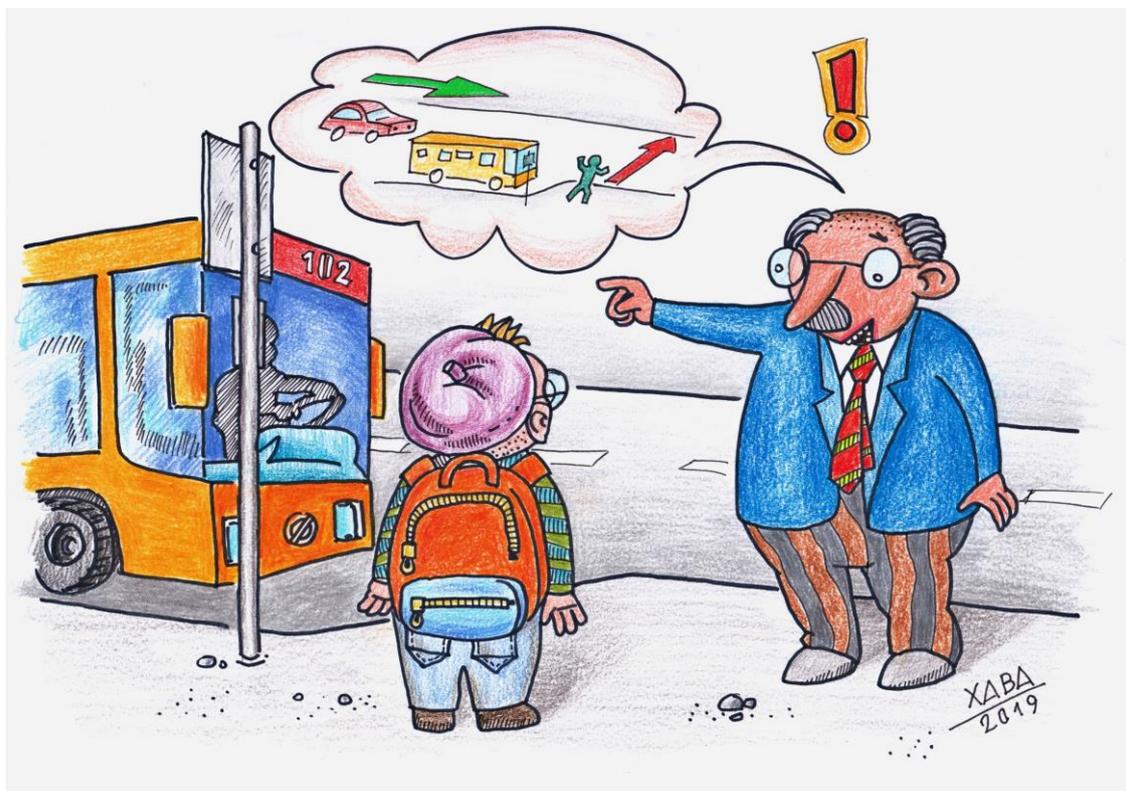

Figure 8. Teaching to understand and follow the well-established rules helps to avoid unexpected risks and to make students thoughtful. Reproduced with permission from Csaba Magyar.

## Acknowledgments

This study was funded by the Content Pedagogy Research Program of the Hungarian Academy of Sciences. The authors thank Csaba Magyar for his imaginative illustration.

## References

1. Boris Korsunsky 2018 Why so series? Phys. Teach. 56, 269 (https://doi.org/10.1119/1.5028252)
2. Solution to the April, 2018 Challenge, Why so series? Phys. Teach. 56, C489 (https://doi.org/10.1119/1.5055312)
3. James Bryant, Walt Jung, Walt Kester, SECTION 7-1 - Passive Components, Editor(s): Walt Jung, Op Amp Applications Handbook, Newnes, 2005, Pages 609-628, ISBN 9780750678445, (https://doi.org/10.1016/B978-075067844-5/50149-1) Open access version (https://www.analog.com/media/en/training-seminars/design-handbooks/Op-Amp-Applications/Section7.pdf )
4. Electrical termination. Wikipedia (https://en.wikipedia.org/wiki/Electrical_termination)
5. Dual Impedance Digital Multimeters, Fluke (https://www.fluke.com/en-us/learn/blog/digital-multimeters/dual-impedance-digital-multimeters)
6. Gingl Z, Mingesz R, Makan G and Mellar J 2019 Driving with Arduino? Keep the lane! Phys. Educ. 54 025010
7. Gingl Z, Makan G, Mellár J, Vadai G and Mingesz R 2018 Phonocardiography, photoplethysmography with simple Arduino setups to support interdisciplinary STEM education, figshare (https://doi.org/10.6084/m9.figshare.7308356.v1)
8. Stojilovic N and Isaacs D E 2018 Resistance of a digital voltmeter: teaching creative thinking through an inquiry-based lab Phys. Educ. 53 053005
9. Makan G, Mingesz R and Gingl Z 2019 How accurate is an Arduino Ohmmeter? Phys. Educ. 54 033001